\documentclass[twocolumn]{aastex701} 

\usepackage{multirow}

\usepackage{graphics,epsf}
\usepackage[utf8]{inputenc}
\usepackage{amsmath}                % American Mathematical Society package
\usepackage{amsfonts}               % American Mathematical Society fonts
\usepackage{amssymb}                % American Mathematical Society symbol
\usepackage{epsfig}                 % EPS figures
\usepackage{graphicx}               % Required for inserting images
\usepackage{float}
\usepackage{color}
\usepackage{multirow}               % double row table entries

\hypersetup{
    colorlinks=true,
    linkcolor=red,   
    urlcolor=cyan}

\usepackage[colorinlistoftodos]{todonotes}

\newcommand{\kms}{{~\rm km\; s^{-1}}}

\newcommand{\km}{{~\rm km}}
\newcommand{\s}{{~\rm s}}

\newcommand{\g}{{~\rm g}}
\newcommand{\G}{{~\rm G}}
\newcommand{\K}{{~\rm K}}

\newcommand{\mum}{{~\rm \mu m}}

\begin{document}

\title{JWST observations support the jittering-jets explosion mechanism (JJEM) for the core-collapse supernova remnant SNR 0540-69.3}

\author[0000-0002-9444-9460]{Dmitry Shishkin}
\affiliation{Department of Physics, Technion - Israel Institute of Technology, Haifa, 3200003, Israel; s.dmitry@campus.technion.ac.il; soker@physics.technion.ac.il}
\email{s.dmitry@campus.technion.ac.il}

\author[0000-0003-0375-8987]{Noam Soker}
\affiliation{Department of Physics, Technion - Israel Institute of Technology, Haifa, 3200003, Israel; 
s.dmitry@campus.technion.ac.il; soker@physics.technion.ac.il}
\email{soker@physics.technion.ac.il}

\begin{abstract}
We examine published JWST observations of the core-collapse supernova (CCSN) remnant SNR 0540-69.3 and identify a point-symmetric morphology in its inner ejecta. Within the framework of the jittering jets explosion mechanism (JJEM), we interpret this morphology as evidence that the ejecta were shaped by two, and likely three or more, pairs of jets during the explosion process. 
Both visual inspection and a recently developed quantitative symmetry-identification method for astrophysical imaging reveal an approximate rotational symmetry between the northeastern redshifted ejecta and the southwestern blueshifted ejecta. Each side contains clumps (knots) surrounding a previously identified cavity, with the best quantitative correspondence obtained for a rotation of $189^\circ$. We further identify a symmetry center that is offset from the current pulsar position, strengthening an earlier claim for a pulsar kick. 
We interpret the pair of cavities and their surrounding clumpy structures as having been shaped by multiple jet-launching episodes. 
In addition, we identify a pair of opposing nozzles at a large angle to the cavities, which we attribute to another jet pair. 
Guided by the similarities to point-symmetric planetary nebulae shaped by jets and by recent three-dimensional hydrodynamical simulations of the JJEM, we conclude that the inner ejecta were shaped by at least three jet pairs launched by the neutron star after it acquired its  kick velocity, consistent with the JJEM.
\end{abstract}
   

% ==================================
\section{Introduction} 
\label{sec:intro}
% ==================================

The community is far from a consensus on the explosion mechanism of core-collapse supernovae (CCSNe). The two competing and heavily studied CCSN explosion mechanisms are the neutrino-driven mechanism (e.g., \citealt{Akhmetalietal2026, ChenCHetal2026, EggenbergerAndersenetal2026, Giudicietal2026, LuoZhaKajino2026, Murphyetal2026, Neopaneetal2026, Orlando2026, PanLi2026, Paradisoetal2026, Rusakovetal2026, VarmaMuller2026, Wessonetal2026}  for some 2026 papers; reviews by \citealt{Janka2025, Mezzacappa2026}), and the jittering-jets explosion mechanism (JJEM; \citealt{Soker2026SN1987Amulecular, Soker2026SNRJ0450, Soker2026Failed, WangShishkinSoker2025, Soker2026DustJets} for some 2026 papers; review by   \citealt{Soker2025Learning}). 

At present, the morphologies of CCSN remnants (CCSNRs) are the only observable that can robustly distinguish between the two theoretical explosion mechanisms (e.g., \citealt{Soker2024UnivReview}), particularly point-symmetric morphologies which the JJEM predicts for many (but not all) CCSNRs (e.g., \citealt{AkashiSoker2026a, AkashiSoker2026BG11, Braudoetal2025, Braudoetal2026}), and the neutrino-driven mechanism has no explanation for. 
 
The first CCSNR with a clear point-symmetric morphology attributed to the JJEM is SNR 0540-69.3. \cite{Soker2022SNR0540} identified the point symmetric structure in Doppler-shift maps that  
\citet{Larssonetal2021} made from their slit spectroscopy observations. This structure is in a plane along the line of sight. \cite{Soker2022SNR0540} also identified a main jet axis in HST images from \cite{Morseetal2006}. \cite{SokerShishkinW49B} compared the rings seen in the images from \cite{Morseetal2006} with circum-jet rings in planetary nebulae and the cluster of galaxies Cygnus A, and, based on these similarities, strengthened the claim for the main jet axis of SNR 0540-69.3. 

Several observational studies have explored various properties of SNR 0540-69.3 (e.g., \citealt{Mathewsonetal1980} discovery; \citealt{Parketal2010, Brantsegetal2014, Lundqvistetal2022, Tenhuetal2025}). Here, we study only the morphology of SNR 0540-69.3 because it is the only property that decisively decides between the two theoretical explosion mechanisms.  

In this study, we argue that the very recent JWST observations of SNR 0540-69.3 in the Large Magellanic Cloud by \cite{Larssonetal2026} present further support for the claim that jets shaped and powered the ejecta of SNR 0540-69.3, hence solidifying the JJEM for this CCSNR. 

% =========================
\section{Clumpy but point-symmetric}
\label{sec:Clumpy}
% =========================

\cite{Larssonetal2026} identified two cavities in the inner ejecta in a bipolar structure, i.e., opposite and touching each other. They take the pulsar to be between the cavities, and from that deduce its natal kick velocity to be $\simeq 300 \km \s^{-1}$ away from the observer. The bubbles are surrounded by a clumpy, filamentary ejecta. \cite{Larssonetal2026} argue that the explosion itself imprints the cavity structure. We accept this argument. The structure of two opposite cavities is very common in cooling flow clusters (e.g., \citealt{McNamaraNulsen2007, Tsaietal2026}), where they are also termed X-ray bubbles. In cooling flow clusters, jets are known to inflate them, as indicated by radio emission. There are some similarities between the morphologies of some jet-shaped X-ray bubbles in cooling flow clusters of galaxies and CCSNRs, suggesting that CCSNRs are also shaped by jets \citep{Soker2024CF}. 
Therefore, we take the structure of two opposite bubbles as inferred by \cite{Larssonetal2026}, to indicate shaping by jets. Moreover, if we adopt the claim for the pulsar being between the bubbles, we conclude, in the framework of the JJEM, that the neutron star (NS) launched the jets that shaped the inner ejecta after it acquired its natal kick. 
Two cavities alone possess axial symmetry, meaning full symmetry around an axis through their centers. Here, we reveal that the bubbles and their surroundings possess a point-symmetric, but not axial-symmetric, structure, indicating shaping by two or more pairs of jets, as the JJEM predicts to occur in some cases. 

\begin{figure*}
\begin{center}
\vspace{-0.0cm}
\includegraphics[trim=0.0cm 24.2cm 0.0cm 0.0cm ,clip, angle=0, scale=0.85]{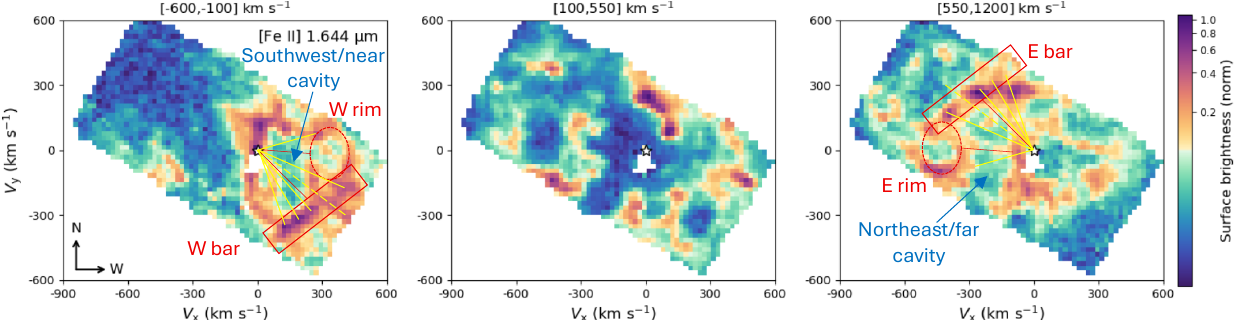} 
\vspace{-0.0cm}
\caption{A figure adapted from \cite{Larssonetal2026} of images of SNR 0540-69.3 in [Fe \textsc{ii}] $1.644 \mum$ emission in three Doppler shift intervals, from left to right: $[-600,- 100] \km \s^{-1}$ $[100,500] \km \s^{-1}$, and $[550,1200] \km \s^{-1}$.  The white star marks the pulsar's position. The white region just south of the pulsar was masked out due to contamination by a star. We marked the two cavities identified by \cite{Larssonetal2026}. On the left panel (blue-shifted), we draw the W-rim, W-bar and connected each to the pulsar with a thin red line. We also connect six knots (bright clumps) to the center with six yellow lines. We then rotate this structure by $180^\circ$ and placed on the right panel (red-shifted part). Five yellow lines point at five knots (clumps) in the redshifted part, while one (dashed line) does not. The E rim is slightly farther from the center. The bar does not match the distance to the clump in the north on the redshifted side. Despite these small discrepancies, the overall morphology is point-symmetric.       
}
\label{Fig:Rims}
\end{center}
\end{figure*}

Figure \ref{Fig:Rims} is from \cite{Larssonetal2026}, showing images in three Doppler-shift ranges as indicated. The left panel is the blue-shifted material, implying the near side of the SNR, while the right panel shows the red-shifted (far) material of the inner ejecta. The faint regions surrounded by the bright material are the cavities, as we mark on these figures. We identify more structural features in these images. In the left panel, the near side, we mark the bright region in the southwest and term it the west bar (W bar). In the west, we identify a bright elliptical rim surrounding a faint zone, which we term the west rim (W rim). We also draw six yellow lines from the center to bright clumps (knots).     
We rotated this entire structure by $180^\circ$, and placed it on the right panel (far side). We find that the east bar is located at a similar distance from the center as the west bar. The east rim is opposite the west rim, but it is somewhat farther: its center is 1.3 times as far from the center that \cite{Larssonetal2026} mark as the west rim. We find 5 bright clumps in the east, in the opposite direction to 5 bright clumps in the west (solid yellow lines), while in one direction of clumps in the west, there is no counterpart in the east (dashed yellow line). Not all clumps are at the same distance. 
% FFFFFFFFFFFFFFFFFFFFFFFFFFFFFFFFF
% FFFFFFFFFFFFFFFFFFFFFFFFFFFFFFFFF

We consider the west/east bar pair and rim pair, with the clumps that compose them, to form a robust point-symmetric morphology. It is not perfect, and cannot be. Several processes may lead to departures from pure point-symmetric morphologies. In the inner ejecta that does not interact with an ambient medium (e.g., \citealt{SokerShishkin2025Vela}), the main effects are that two opposite jets in a pair might be unequal in their properties, like energy and opening angle (e.g., \citealt{Soker2024CounterJet}), the instabilities that occur in the jet-medium interaction as three-dimensional (3D) simulations show (e.g., \citealt{Braudoetal2025}), and that the post-kick jets that the NS launches are displaced with respect to the pre-kick jets (e.g., \citealt{SokerShishkin2025Vela}). 

% =========================
\section{Quantitative symmetry analysis }
\label{sec:Quantitative}
% =========================

% ==================================
\subsection{Symmetry identifier}
\label{subsec:clumpy_symm}
% ==================================

To quantify and further characterize the point-symmetric morphology, we apply the symmetry identification method of \cite{ShishkinMichaelis2026} to the blue- and red-shifted components of the inner ejecta.

The transformation information (TI) method quantifies the similarity between image-intensity distributions using a Kullback-Leibler cross-entropy measure. By continuously (and discretely) transforming one image relative to another, the TI score is evaluated as a function of the transformation parameter (here: rotation angle). Local minima in this score correspond to configurations with maximal distributional overlap. For self-comparisons, these minima identify symmetry axes. For comparisons between distinct images, they instead highlight localized correspondences, indicating similar spatial intensity distributions between the two datasets (images).

We use the G140H/F100LP stage 3 data product (program \#4712) downloaded from the Mikulski Archive for Space Telescopes (MAST) and do not apply any additional post-processing.
We recreate the stacked velocity images of \cite{Larssonetal2026}, adopting a narrower velocity interval for the red-shifted image to ensure that both the blue- and red-shifted stacks contain the same number of valid frames (10). 
We then apply a rotational transformation to the [700,1150] $\kms$ image (hereafter the red-shifted image) and compute the TI score as a function of rotation angle relative to the [-600,-100] $\kms$ image (hereafter the blue-shifted image).

\begin{figure*}
\begin{center}
\includegraphics[trim=0cm 0cm 0cm 0cm ,clip, width=\textwidth]{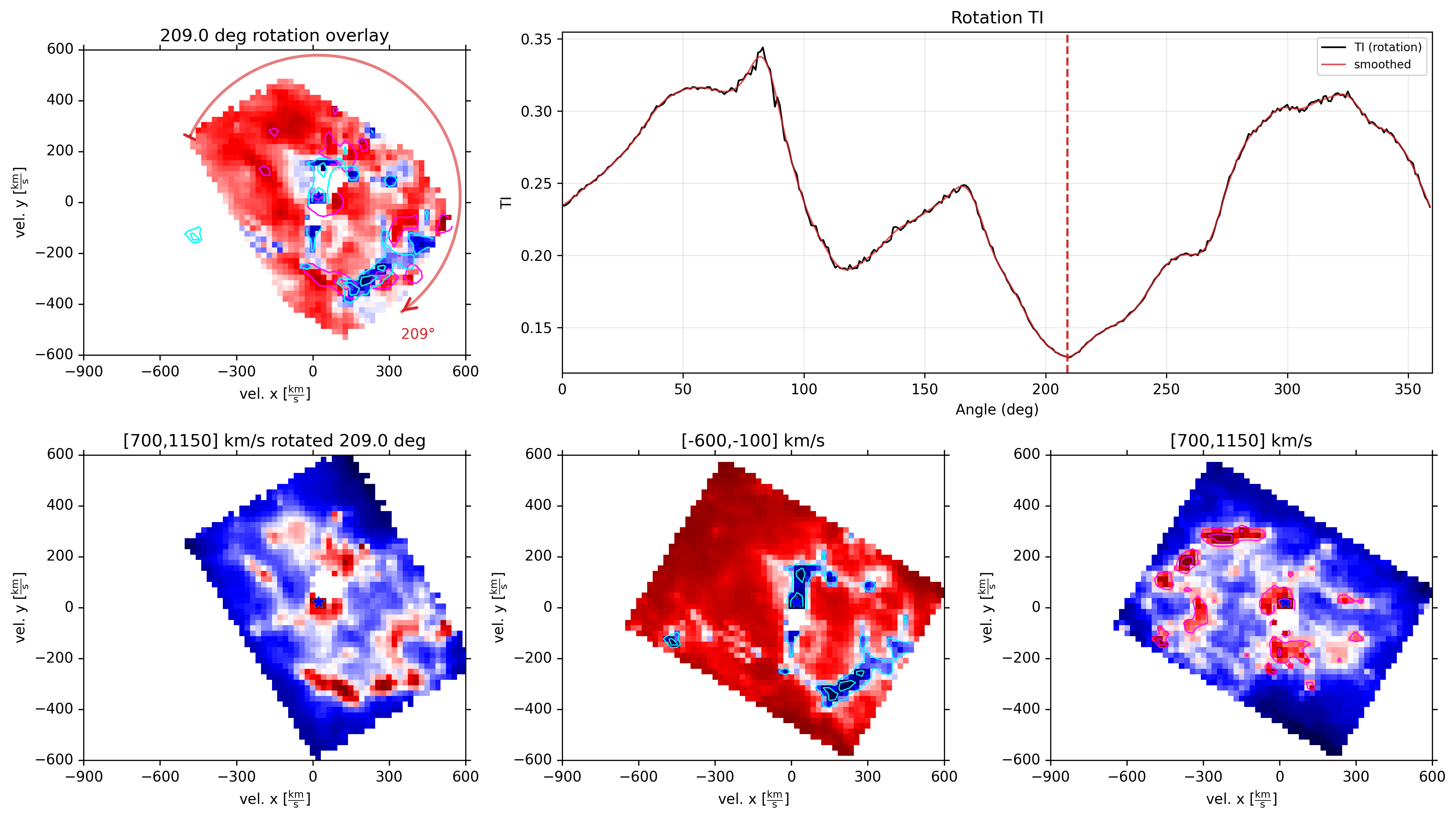} 
\caption{Application of the symmetry-identification method to the blue-shifted stacked image and the red-shifted stacked image. The lower-right panel is the red-shifted image of the surroundings of the northeast/far cavity, but at a slightly different velocity range than in Figure \ref{Fig:Rims}. The lower-middle panel is the blue-shifted/near cavity in the same velocity range as in Figure \ref{Fig:Rims}. The lower-left panel shows the rotation of the redshifted image by $209^\circ$ clockwise. 
The upper-left panel overlays the left and middle lower panels. 
The top-right panel shows the TI score as a function of rotation angle between the transformed blue-shifted image ([700,1150] $\kms$) and the red-shifted image ([-600,-100] $\kms$). The global minimum at $\theta_{\rm TI,min}^{\rm NS}=209^\circ$ corresponds to the alignment of the western and eastern bars (W and E bars, see Figure~\ref{Fig:Rims}) when the blue-shifted image is rotated about the current NS position (blue star).
We denote contour lines in both images to better highlight the brightest regions and demonstrate when they overlap in the overlay image (top-left panel).}
\label{Fig:symm_NS}
\end{center}
\end{figure*}

We present the blue-shifted and red-shifted images in the bottom-center and bottom-right panels of Figure~\ref{Fig:symm_NS}, respectively. 
The upper panel presents the TI score as a function of rotation angle, with the global minimum occurring at $\theta_{\rm TI,min}^{\rm NS}=209^\circ$. The bottom-left panel shows the blue-shifted image rotated by this angle, while the upper-left panel displays the overlay of the rotated blue-shifted image and the non-rotated red-shifted image.
% FFFFFFFFFFFFFFFFFFFFFFFFFFFFFFFFF
% FFFFFFFFFFFFFFFFFFFFFFFFFFFFFFFFF

The global minimum at $\theta_{\rm TI,min}^{\rm NS}=209^\circ$, corresponding to the best match between the red-shifted and blue-shifted images, aligns the most prominent outer structures with one another: the western and eastern bars (W and E bars in Figure~\ref{Fig:Rims}). An exact correspondence between the smaller-scale clumpy substructures cannot be achieved through a simple rotation about the current NS position. This is expected, as the clumps are neither equidistant from the NS nor symmetrically distributed about the axis connecting them.

% ==================================
\subsection{An alternative center}
\label{subsec:clumpy_center}
% ==================================

The minima of the TI score identify candidate symmetry axes, while the minimum TI value provides a quantitative measure of the degree of symmetry. We therefore perform a grid search over the pixel region surrounding the current NS location and identify the center that minimizes the global TI score, corresponding to the location with the highest degree of point symmetry.

\begin{figure*}
\begin{center}
\includegraphics[trim=0cm 0cm 0cm 0cm ,clip, width=\textwidth]{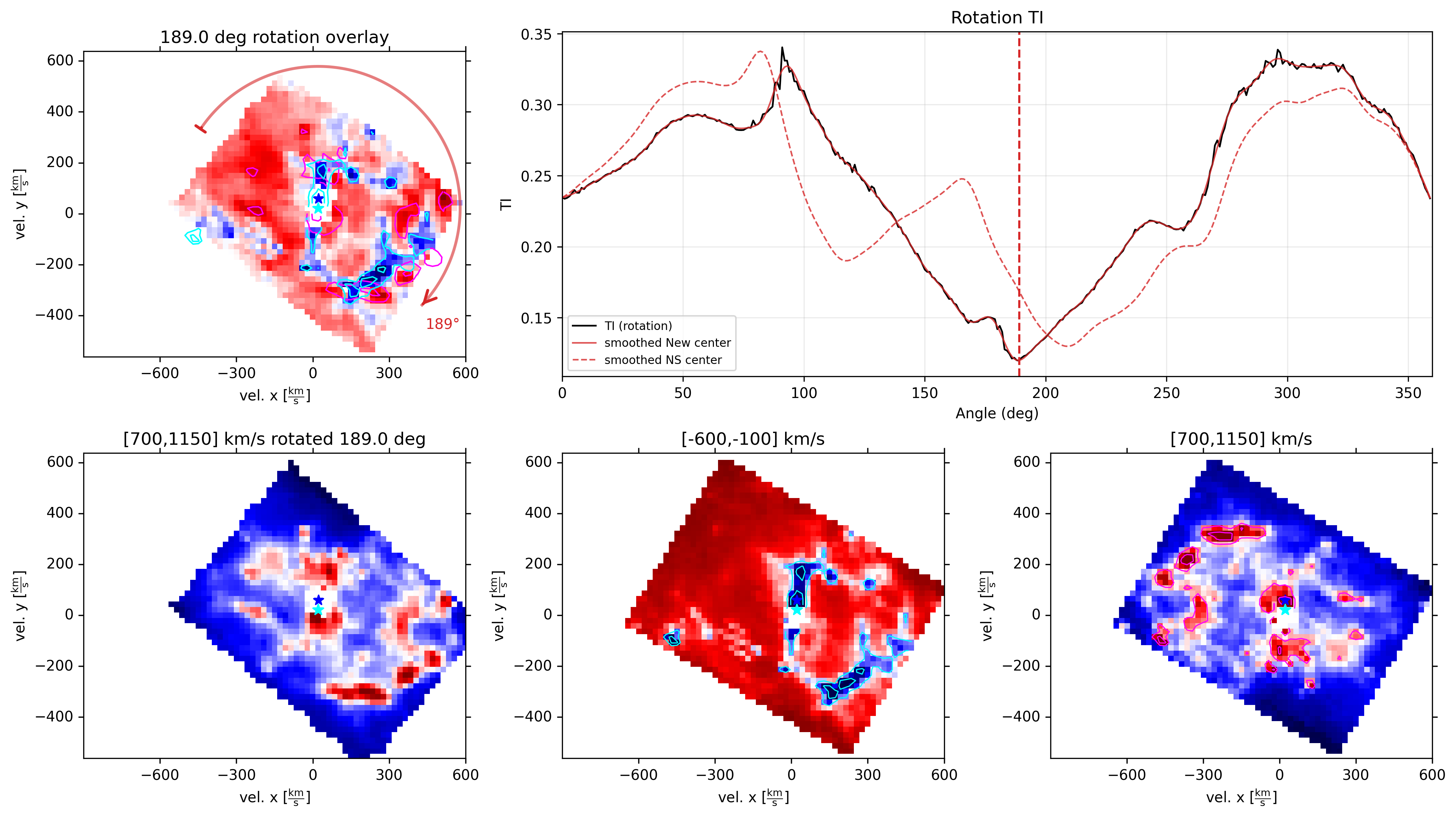} 
\caption{Similar to Figure~\ref{Fig:symm_NS}, but adopting the optimized center location (cyan star) obtained from the TI grid search, just south of the pulsar (black star), and for a rotational symmetry of $\theta_{\rm TI,min}^{\rm new}=189^\circ$. The new location produces a better correspondence between both the outer bars and the inner structures that are not masked. The clumpy emission within the eastern and western bars exhibits a partial spatial overlap. For comparison, we show the TI curve from Figure~\ref{Fig:symm_NS} as a dashed line, illustrating both the lower minimum TI score (0.12 versus 0.13) and the shift in the rotation angle at which the optimum alignment is achieved.}
\label{Fig:symm_New}
\end{center}
\end{figure*}

Figure~\ref{Fig:symm_New} repeats the analysis presented in Figure~\ref{Fig:symm_NS}, but using this optimized center. Compared to the current NS position, the new center yields a lower global TI score and a more consistent alignment of both the large-scale bars and the inner clumpy substructure. In particular, the clumpy emission within the eastern and western bars exhibits a noticeably improved overlap, although the correspondence remains imperfect.
% FFFFFFFFFFFFFFFFFFFFFFFFFFFFFFFFF
% FFFFFFFFFFFFFFFFFFFFFFFFFFFFFFFFF

The optimized center is obtained by minimizing the TI score under rotational transformations alone (Section~\ref{subsec:clumpy_symm}). While this approach improves the identification of point-symmetric structures, it is unlikely to capture the full complexity of the ejecta morphology. More complex transformations (e.g., including scaling) may provide a better description of the symmetry, particularly the potentially asymmetrical nature of early jet pairs.

For Figure~\ref{Fig:symm_New}, we also adopt the optimized center as the velocity center, resulting in a slight change to the displayed field of view.

We list several caveats and limitations of this method here.
First, the relatively low spatial resolution of the observations (approximately $30\times60$ pixels) makes the sub-pixel interpolation scheme used during image rotation non-negligible. We retain the linear interpolation as in \cite{ShishkinMichaelis2026}.
Second, we do not apply any additional point-source removal other than the masking of the same pixels covering the star near the NS location as in \cite{Larssonetal2026}. Since these point sources are comparable in size to the clumps of interest, they may influence the absolute TI values. However, inspection of the image overlays at the global TI minimum indicates that these sources do not systematically coincide with the brightest overlapping structures and are therefore unlikely to dominate the inferred symmetry.

We attribute the point symmetric structure of the two cavities and their surroundings to two or more pairs of jets that the NS launched after it acquired its natal kick. The directions of the jittering jets in the JJEM are largely random. Therefore, the inner ejecta symmetry axes might be uncorrelated with the symmetry axes of the outer ejecta that previous studies identified and attributed to jets in the framework of the JJEM (e.g., \citealt{Soker2022SNR0540, SokerShishkinW49B}). 

% ==================================
\section{Jet-shaped nozzles} 
\label{sec:Nozzles}
% ==================================

\begin{figure*}
\begin{center}
\vspace{-0.0cm}
\includegraphics[trim=0.0cm 7.0cm 0.0cm 0.0cm ,clip, angle=0, scale=0.85]{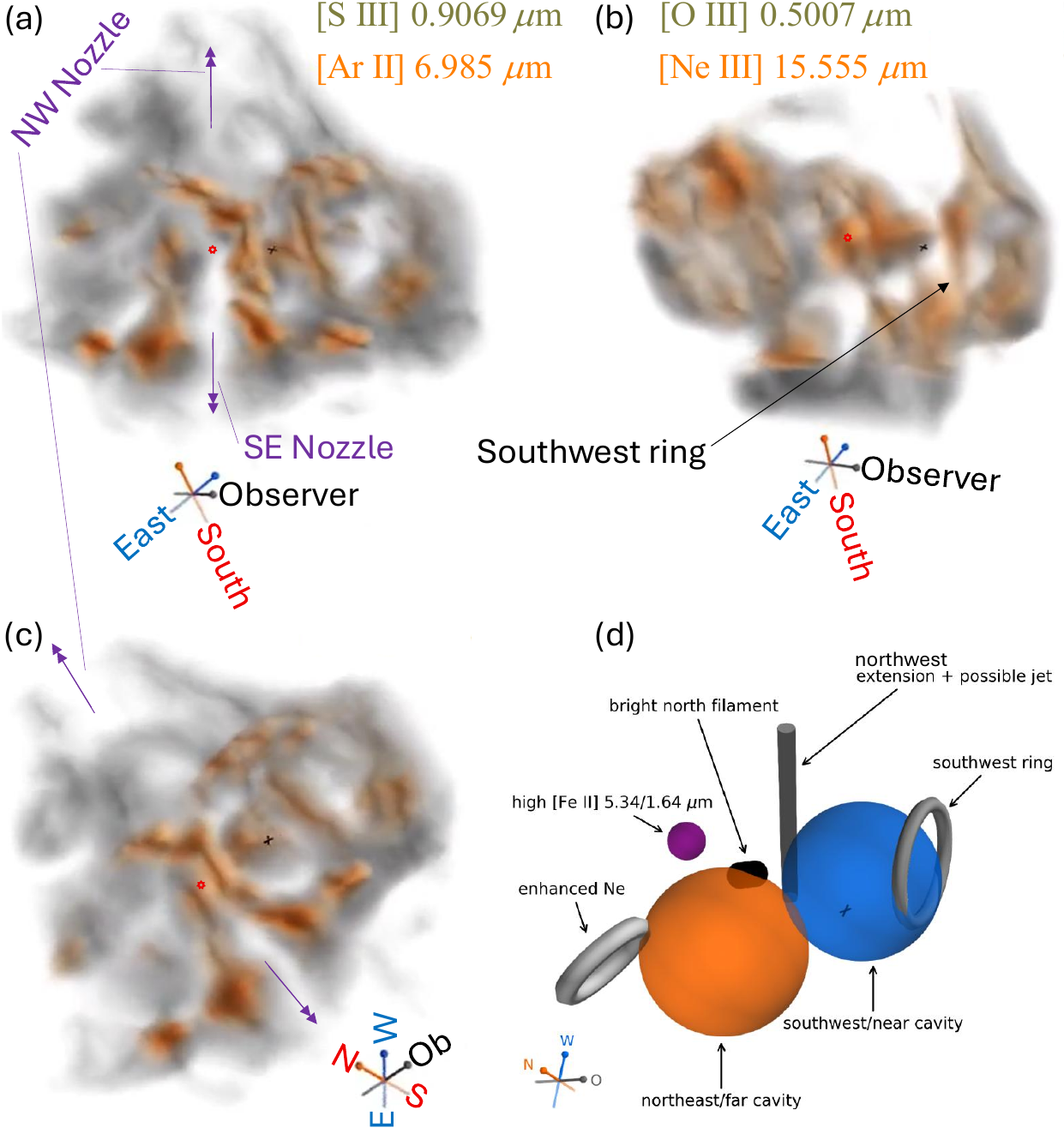} 
\vspace{-0.0cm}
\caption{Images of SNR 0540-69.3 adapted from \cite{Larssonetal2026}, with our marks of the nozzles by two double-headed arrows. We attribute the nozzles to a pair of jets at the end of the explosion process. The axis of the nozzles is at a large angle to the axis connecting the centers of the two spheres, which signify two cavities. We attribute the two cavities to another pair, or more, of jets at the end of the explosion process (see text). The red circle is the point shifted by $-300 \km \s^{-1}$ from the center of explosion, as marked by the black cross in \cite{Larssonetal2026}. Panels (a) - (c) compare the 3D volume structure of different emission lines. The color code of panel (c) is the same as that of panel (a). We took the images from the animation of the online version of the paper.     
(d) A schematic view of the main components of the 3D morphology that \cite{Larssonetal2026} identified (we replaced their `northeast extension' with a `northwest extension' that fits better). The length of each compass axis is $200 \km \s^{-1}$, and the center of explosion that \cite{Larssonetal2026} assumed is marked by a black cross. The two spheres that approximate the cavity have diameters of $600 \km \s^{-1}$.
}
\label{Fig:Nozzles}
\end{center}
\end{figure*}

\cite{Larssonetal2026} present a 3D image of [S \textsc{iii}] $0.9069 \mum$ and [Ar \textsc{ii}] $6.985 \mum$ emission, with an animation in the online material that changes the viewing angle. That image shows hints of two opposite nozzles. In Figure \ref{Fig:Nozzles}, we present two images from their animation at two different viewing angles (panels a and c); we also present one image of different emission lines at a different viewing angle (panel b). Panels (a) and (c) reveal two opposite nozzles: the NW and the SE nozzles. 3D hydrodynamical simulations of the JJEM obtain such nozzles (e.g., \citealt{Braudoetal2026, AkashiSoker2026a}), and similar nozzles that are attributed to jets exist in many planetary nebulae, e.g., Hen 2-115 (PN G321.3+02.8; in \citealt{SahaiTrauger1998}), M1-6 (PNG 211.2-03.5), and Hen 2-86 (PNG 300.7-02.0; in \citealt{Sahaietal2011}). Based on these, we attribute the nozzles of SNR 0540-69.3 to jet shaping. 
% FFFFFFFFFFFFFFFFFFFFFFFFFFFFFFFFF
% FFFFFFFFFFFFFFFFFFFFFFFFFFFFFFFFF

In Sections \ref{sec:Clumpy} and \ref{sec:Quantitative}, we argued that a pair or more of jets inflated the bubbles during the explosion. The rings, one at the surface of each cavity away from the center that \cite{Larssonetal2026} identify (panel d of Figure \ref{Fig:Nozzles}), suggest that the jets propagated further out than the cavities, as jets can inflate circumjet rings, as observations in different types of astrophysical objects (e.g., \citealt{SokerShishkinW49B}) and simulations of the JJEM (e.g., \citealt{AkashiSoker2026a, AkashiSoker2026BG11}) show. The cavities do not seem to be part of a torus, but rather two cavities, as observed in other astrophysical types of objects that are inflated by jets. The nozzle axis is at a large angle to the cavity axes, possibly almost perpendicular. We suggest that the two axes point at two or more pairs of jets that the NS (pulsar) launched after it acquired its natal kick. The complex structure of the cavities, with the bar and rim (Figure \ref{Fig:Rims}), suggests that either more than one pair of jets inflated the cavities, or that these jets precessed. The pulsar wind nebula also shaped the inner ejecta. However, 3D simulations of pulsar wind nebulae (e.g., \citealt{Porthetal2014, Porthetal2017, Olmietal2016}) usually form structure along the polar axis and in the equatorial plane. They do not form a structure of two inclined axes, as we inferred for the inner ejecta of SNR 0540-69.3.

% ==================================
\section{Discussion and Summary} 
\label{sec:Summary}
% ==================================

We analyzed recently published JWST observations of SNR 0540-69.3. 
By eye inspection, we qualitatively identified a point symmetric structure in the blue and red shifted maps from \cite{Larssonetal2026}, as we showed in Figure \ref{Fig:Rims}. We then quantified this symmetry using the quantitative symmetry-identification pipeline developed by \cite{ShishkinMichaelis2026} for astrophysical imaging analysis. This method is based on Transformation Information (TI), which is a measure of self-similarity under geometric transformations. This allows us to find a better center for the rotational symmetry, as marked by a cyan star in Figure \ref{Fig:symm_New}, which gives a better quantitative fit to the rotational symmetry than when the center is the pulsar (Figure \ref{Fig:symm_NS}). Also, the rotation angle with the new center is $189^\circ$, much closer to $180^\circ$ than the rotation angle of $209^\circ$ with the pulsar at the center.  

As we mark in Figure \ref{Fig:Rims}, the two-sided point-symmetric structure we analyze in Figures \ref{Fig:Rims} - \ref{Fig:symm_New} is the structure around the two opposite cavities that \cite{Larssonetal2026} identified. In cooling flow clusters of galaxies, pairs of jets inflate pairs of cavities; there are similarities in some jet-shaped morphological features in cooling flow clusters and CCSNRs \citep{Soker2024CF}. We attribute the shaping of the two cavities of SNR 0540-69.3 to one or more pairs of opposite jets. 
A key property is that this two-sided symmetry is not an axial symmetry, as evident from the rims and bars' structural features that we defined in Figure \ref{Fig:Rims}. This suggests a precessing pair of jets, or more likely, two or more pairs of jets that shaped the inner ejecta: the cavities and their surroundings. We do not find the cavities to be part of a large torus. 

We further identified two opposite nozzles with an axis at a large angle (close to being perpendicular) to the axis connecting the centers of the two cavities. We use 3D images from \cite{Larssonetal2026} to identify the pair of nozzles, as we marked in Figure \ref{Fig:Nozzles}. The NW nozzle seems to coincides with the northwest extension that \cite{Larssonetal2026} marked (lower right panel of Figure \ref{Fig:Nozzles}). SNR J0450.4-7050 has a similar pair of nozzles in the inner ejecta \citep{Soker2026SNRJ0450}; \cite{Braudoetal2026} demonstrate that pairs of jets can form this morphology in the inner ejecta. 

The method of identifying point-symmetric morphologies, such as multipolar structures, and, from that, deducing that jets shaped the ejecta is a common practice in the study of planetary nebulae (e.g., \citealt{Morris1987, Soker1990AJ, Guerreroetal1998,  Guerreoetal2021, SahaiTrauger1998, Sahaietal2005, Boffinetal2012, RechyGarciaetal2019, Tafoyaetal2019, GarciaSeguraetal2020, GarciaSeguraetal2021, Clairmontetal2022, RechyGarciaetal2020, Clairmontetal2022, MoragaBaezetal2023,  Derlopaetal2024, Mirandaetal2024, Sahaietal2024}). The morphological similarities between many CCSNRs, including SNR 0540-69.3 that we analyzed here, and many jet-shaped planetary nebulae are no coincidence. It is a strong clue that multiple pairs of jets are responsible for CCSN explosions. 

\cite{Soker2022SNR0540} already identified a point symmetric structure in SNR 0540-69.3, but in a larger area and in Doppler-shift maps published by \cite{Larssonetal2021}; this structure is in a plane along the line of sight. \cite{Soker2022SNR0540} attributed this structure to the JJEM. In addition, SNR 0540-69.3 has a main jet axis on the plane of the sky, as  \cite{Soker2022SNR0540} identified in HST images from \cite{Morseetal2006}, and circum-jet rings that \cite{SokerShishkinW49B} studied. The point-symmetric structure we identified here is closer to the center. The outer and inner point-symmetric structures might not be correlated, as it is possible that the inner point-symmetric structure is due to later jittering jets, which the NS launched after it acquired its natal kick. 

Overall, we consider both the qualitative and quantitative evidence for a point-symmetric morphology in the inner ejecta of SNR 0540-69.3 to be robust. We interpret this morphology as evidence that the inner ejecta were shaped by at least two, and more likely three or more, pairs of jets. This interpretation is motivated by the point-symmetric morphologies observed in numerous planetary nebulae that are inferred to have been shaped by jets, as well as by recent three-dimensional hydrodynamical simulations of the JJEM (Section~\ref{sec:intro}).
Our results support the JJEM and are difficult to reconcile with explosion models that rely solely on neutrino-driven mechanisms (neutrino heating; delayed neutrino).

% ===================================================
\section*{Acknowledgements}
% ===================================================

This work is based on observations made with the NASA/ESA/CSA James Webb Space Telescope. The data were obtained from the Mikulski Archive for Space Telescopes at the Space Telescope Science Institute, which is operated by the Association of Universities for Research in Astronomy, Inc., under NASA contract NAS 5-03127 for JWST. These observations are associated with program \#4712.

% https://journals.aas.org/facility-keywords/
\facility{JWST}

% https://journals.aas.org/policy-statement-on-software/
\software{NumPy \citep{harris2020array},
          OpenCV \citep{opencv_library},
          Matplotlib \citep{Hunter:2007},
          symmetries\_snr (\citealt{ShishkinMichaelis2026}, \href{https://github.com/AmirGoshenMichaelis/symmetries\_snr}{GitHub: symmetries\_snr})
          }

\end{document}